\documentclass{emulateapj}
\usepackage{apjfonts}
\usepackage{natbib}
\usepackage{graphicx}
\usepackage{amssymb,amsmath}
\usepackage{xcolor}
\definecolor{ao}{rgb}{1.0,0.03,0}
\definecolor{ap}{rgb}{0, 0.18,0.39}
\usepackage[breaklinks,colorlinks,linkcolor=ao,citecolor=ap]{hyperref}
\shorttitle{A global ILC approach using CMB covariance matrix}

\begin{document}

\title{A Global ILC Approach in Pixel Space over Large Angular Scales of the Sky using CMB Covariance Matrix} 

\author{Vipin Sudevan\altaffilmark{1},  Rajib Saha\altaffilmark{1}} 

\altaffiltext{1}{Physics Department, Indian Institute of Science 
Education and Research Bhopal,  Bhopal, M.P, 462066, India.}

\begin{abstract}
We propose a new internal linear combination (ILC) method in the pixel space, 
applicable on large angular scales of the sky, to estimate a foreground minimized 
Cosmic Microwave Background (CMB) temperature anisotropy map by incorporating 
prior knowledge about the theoretical CMB covariance 
matrix. Usual ILC method in pixel space, on the contrary, does not use any information about the 
underlying CMB covariance matrix. 
The new  approach complements the usual pixel space  ILC technique 
specifically at low multipole region, using global information available from theoretical CMB covariance matrix
as well as from the data. Since we apply our method over the large scale on the
sky containing low multipoles we perform foreground minimization globally. %We perform foreground 
We apply our methods  on low resolution Planck and WMAP foreground 
contaminated CMB maps and validate the methodology by performing 
detailed Monte-Carlo simulations. Our cleaned CMB map and its power spectrum  
have significantly less error than  those obtained  following usual ILC technique
at low resolution that does not use CMB covariance information. Another very important 
advantage of our method is that the cleaned power spectrum does not have any negative 
bias at the low multipoles because of effective suppression of CMB-foreground chance 
correlations on large angular scales of the sky. Our cleaned CMB map  and its power spectrum  
match well with those estimated by other research groups. 
\end{abstract}

\keywords{cosmic background radiation --- cosmology: observations --- diffuse radiation}
\maketitle

\section{Introduction}
For reconstruction of Cosmic Microwave Background (CMB) signal from multi-frequency observations 
an important method is  Internal-Linear-Combination (ILC)~\citep{Tegmark96, Tegmark2003, Bennett2003,
Eriksen2004, Saha2006, Hinshaw_07}. To obtain a foreground minimized CMB map the ILC method requires neither to 
explicitly model the frequency spectra of individual foreground components, nor does it require to model 
the foreground amplitudes (at some reference frequency) in terms of so called foreground templates.  
The only assumption one makes related to foregrounds is that each of them has a frequency spectrum that is 
different from the frequency spectrum of CMB component, which is assumed to be that of black-body in nature~\citep{Mather1994, Fixen1996}.  The 
basic idea behind the ILC method is to linearly superpose the available foreground contaminated CMB maps  using 
certain amplitude terms,  a set of weights, to estimate a foreground minimized CMB map. The weights are 
obtained by minimizing the variance of the cleaned map and can be computed analytically by using a simple 
formula.  In spite of being simple to design and yet a powerful  technique to reconstruct a cleaned CMB 
map we see that it is necessary -- for a few important reasons -- to investigate performance of the usual ILC method in some 
hitherto unexplored cases.  First, while estimating the weights the usual ILC method in pixel space 
does not take into account the covariance structure of the CMB maps. In other words, it does not use 
the fact that the final cleaned map, if perfectly cleaned of all foregrounds and detector noise is negligible, 
should have a covariance structure consistent with the underlying theoretical model. Secondly, some of the 
maximum likelihood methods~\citep{Eriksen2007, Eriksen2008, Eriksen2008a, Gold2011, PlanckFg2016, PlanckCMB2016} 
for component separation however use CMB and detector noise covariance matrices to reconstruct 
CMB and all foreground components. It is therefore natural to ask a question {\it can we generalize usual 
ILC method in pixel space to incorporate  CMB covariance information also?}   

In the present work we seek to find a solution to the above problem and generalize the pixel space  ILC 
method taking into account prior information of the theoretical covariance matrix of the CMB maps. Therefore, instead of 
minimizing simple variance of the cleaned map we propose to estimate the weights by minimizing the reduced 
variance of the cleaned map, the reduced variance being defined  by the CMB covariance weighted 
variance of the cleaned map, which is explained in Section~\ref{formalism}. Since storage space
into the computer disks  of such full pixel-space covariance matrix 
increases rapidly with the HEALPix pixel resolution parameter $N_{side}$ ($\sim N^4_{side})$, in the current work 
we use low pixel resolution maps.  Further to focus largely on the low multipoles  we smooth the 
input $N_{side} =16$ maps by a Gaussian window function of FWHM $9^\circ$. 
At this smoothing the input maps contain approximately $2.5$ pixels per beam width, which implies these maps are
properly band-width limited.  The larger beam smoothing also reduces  detector noise contributions at different 
pixels.

Our method at low resolution bears an interesting complementarity in its approach when compared with the usual 
pixel space ILC method at high resolution, that do not use the CMB covariance matrix.  Since the level of 
foreground contamination, and their spectral properties  vary with the sky positions, in a high resolution 
analysis of usual ILC method one performs foreground cleaning individually over several smaller regions of the 
sky, in such a way that the foreground spectral properties and level of foreground contaminations in each 
region remains approximately constant. Because of low pixel resolution (and large smoothing on the low pixel 
resolution maps) of this work we chose either to perform foreground removal over the entire sky or by dividing 
the sky into  small number of regions. In the second approach we divide the sky into two regions and clean them 
individually in a total of two iterations. Our aim is to  use as much large sky fraction as possible during 
foreground removal and information about CMB theoretical covariance matrix from the corresponding large fraction, 
so that our method  becomes a global method of foreground minimization. Thus our method may be seen as dual to 
usual high resolution ILC method, wherein the former uses global information from the covariance matrix and the 
data to estimate the foreground minimized CMB map and the later relies upon the local information of foregrounds 
properties. 

By performing detailed Monte-Carlo simulations we find that the new ILC method of this work  has significantly less 
reconstruction  errors in cleaned maps and power spectrum than the usual ILC method in pixel space over 
large angular scales of the sky. The cleaned power spectrum of our method does not have a negative bias 
at the low multipole region that is present in usual ILC method and is caused by a chance correlations between 
CMB and foreground components on a particular realization of CMB sky. 

The subject of component separation in the context of CMB is very rich.~\cite{Bunn1994, Bouchet1999} propose 
a Wiener filter approach.~\cite{Saha2008} discuss in detail 
bias issues in CMB angular power spectrum for harmonic space ILC approach.~\cite{Saha2016} apply an ILC 
technique to jointly estimate CMB and foreground components for Stokes Q polarization in presence of 
varying spectral index of synchrotron component. Iterative harmonic space ILC algorithm was applied on
high resolution Planck and WMAP data, and one of its limitations arising due to foreground leakage 
was first discovered and remedied by~\cite{Sudevan2017}.~\cite{Delabrouille2012} and \cite{Delabrouille2013}
implement a needlet space ILC algorithm to incorporate localization of foreground emissions both in 
pixel space and its `Fourier' space. A variant of ILC technique by minimizing a measure of non-Gaussianity 
was implemented on WMAP temperature and Polarization data by~\cite{Saha2011} and~\cite{Purkayastha2017} respectively.   
~\cite{Eriksen2007, Eriksen2008, Eriksen2008a} propose Gibbs sampling for  component separation.~\cite{Gold2011}
use Markov Chain Monte Carlo method to jointly estimate CMB and foregrounds from WMAP data.

We organize our paper as follows. In Section~\ref{formalism} we discuss the formalism of the new method. We 
describe  how to compute the theoretical CMB covariance matrix in Section~\ref{comp_C}  and comment on 
its singular nature in Section~\ref{singC}. In Section~\ref{method} we describe in detail our foreground 
minimization approaches on Planck and WMAP low resolution maps. We discuss the cleaned maps and CMB angular 
power spectra obtained from data  on Section~\ref{Result}. We  validate  our foreground minimization methods by 
performing Monte Carlo simulations in Section~\ref{validity}. In Section~\ref{advantage} we show the advantage 
of the new ILC approach in pixel space over the usual ILC approach for analysis over large angular scales 
on the sky. We investigate the role of CMB-foreground chance correlation in not-so-efficient foreground 
removal by the usual ILC methods at low resolution in Section~\ref{chancecorr} and comment that using the
CMB covariance matrix in our new method, we effectively suppress such chance correlations which leads to 
improved foreground minimization.  Finally we conclude in Section~\ref{Conclusion}.
 
\section{Formalism} 
\label{formalism}
Let we have $n$ full sky foreground contaminated CMB maps,  ${\bf X}_{i}$ at a 
frequency $\nu_i$, with $i = 1, 2, ...., n$  at some  beam and pixel resolution 
in thermodynamic temperature unit. We assume mean temperature corresponding to 
each frequency $\nu_i$ has already been subtracted from each ${\bf X}_{i}$. 
${\bf y}$ represents the cleaned CMB map obtained by linear combination 
of $n$ input maps  ${\bf X}_{i}$, with weight factor $w_{i}$, i.e., 
\begin{eqnarray} 
{\bf y} = \sum_{i=1}^nw_{i} {\bf X}_{i}\, .
\label{cmap}
\end{eqnarray}  
Here each ${\bf X}_i$  and ${\bf y}$ are  $N\times 1$ vectors describing  full sky 
HEALPix\footnote{Hierarchical Equal Area Isolatitude Pixellization of sphere, e.g., see
 \cite{Gorski2005}} map with $N$ pixels for a pixel resolution parameter $N_{side}$ 
($N = 12N^2_{side}$), smoothed  by Gaussian beam of certain FWHM. Instead of
 minimizing cleaned map variance ${\bf y}^T{\bf y}$  like the usual pixel space ILC 
method we propose a more general approach by incorporating the prior information 
about the theoretical CMB covariance matrix. We  minimize, 
\begin{eqnarray}
\sigma^2 = {\bf y}^T{\bf C}^{\dagger}{\bf y}\, ,  
\label{dispersion0}
\end{eqnarray} 
where ${\bf C}$ represents the CMB theoretical covariance matrix which as discussed in 
Section~\ref{singC} may not be  always invertible. ${\bf C}^{\dagger}$ represents Moore-Penrose 
generalized inverse~\citep{Moore1920, Penrose1955} of matrix ${\bf C}$.  Using Eqn.~\ref{cmap}  we 
can write Eqn.~\ref{dispersion0} as  
\begin{eqnarray}
\sigma^2 = {\bf W} {\bf A W}^T \, , %\sum_{i',j'=1}^n w_{i'} A_{i'j'}w_{j'}\, ,  
\label{dispersion1} 
\end{eqnarray} 
where ${\bf W} = \left(w_1, w_2, w_3, ...., w_n\right)$ is a $1\times n$ row 
vector of weight factors of different frequency maps and ${\bf A}$ is an $n\times n$ matrix  
with its elements $A_{ij}$ satisfying
\begin{eqnarray}
A_{ij} = {\bf X}^T_{i}{\bf C}^{\dagger} {\bf X}_{j}\, .
\label{AMatrix} 
\end{eqnarray} 
Since spectral distribution of CMB photons is that of a blackbody to a very good approximation, CMB 
anisotropy in thermodynamic temperature unit is independent on frequency bands. To reconstruct 
CMB anisotropies without introducing any multiplicative bias in its amplitude we constrain 
the weights for all frequency bands to sum to unity, i.e., $\sum_{i=1}^n w_{i} = 1$.  
The choice of weights that minimize the variance given by Eqn.~\ref{dispersion0} is obtained 
following a Lagrange's multiplier approach (e.g., see \cite{Saha2008} and also \cite{Tegmark96, 
Tegmark2003, Saha2006})  
\begin{eqnarray}
{\bf W} = \frac{ {\bf e} {\bf A}^{\dagger}}{ {\bf e} {\bf A}^{\dagger} {\bf e}^T}\, ,
\label{weights}
\end{eqnarray}  
where ${\bf A}^{\dagger}$ represents Moore-Penrose generalized inverse
of matrix ${\bf A}$ and ${\bf e} = \left(1, 1, ..., 1\right)$ is a $1\times n$ row-vector 
representing shape vector of CMB in thermodynamic temperature unit. 

\begin{figure*}
\includegraphics[scale=0.62]{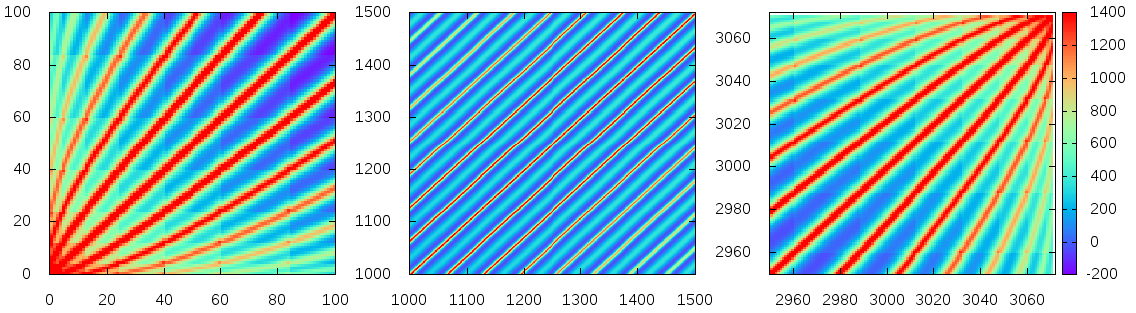}
\caption{Different (square) block matrices taken across the diagonal of the theoretical CMB covariance matrix, ${\bf C}$. The non 
diagonal nature of this matrix is consistent with statistically isotropic CMB. Color map unit is in $\mu K^2$ (thermodynamic). }
\label{covmat} 
\end{figure*}

\section{Computing CMB Covariance Matrix}
\label{comp_C}
To compute elements of the CMB covariance matrix, ${\bf C}$  we assume  principle of statistical 
isotropy of CMB anisotropy. Under this assumption the elements $C_{ij}$ of matrix  ${\bf C}$,  at 
the chosen beam and pixel resolution are given by
\begin{eqnarray}
C_{ij} = \sum_{\ell =2}^{\ell=\ell_{max}}\frac{2\ell+1}{4\pi} C_{\ell}B^2_{\ell} 
\mathcal P_{\ell}(\cos(\gamma_{ij}))P^2_{\ell} \, ,
\label{theory_cov}
\end{eqnarray}  
where $C_{\ell}$ is the fiducial CMB angular power spectrum~\citep{PlanckCosmoParam2016}, 
$B_{\ell}$ represents the  beam transfer function, $\mathcal  P_{\ell}$ denote Legendre
polynomials and $P_{\ell}$ is pixel window function for the given  $N_{side}$ parameter.  
The cosine of the angle $\gamma_{ij}$ is obtained following
\begin{eqnarray}
\cos(\gamma_{ij}) = \cos(\theta_i)\cos(\theta_j) + 
\sin(\theta_i)\sin(\theta_j)\cos(\phi_i - \phi_j) \, ,
\label{gammaij} 
\end{eqnarray}     
where $(\theta_i, \phi_i)$ and $(\theta_j, \phi_j)$ are spherical polar angles respectively of $i$ 
and $j$th pixels of the map. Under the assumption of statistical isotropy ${\bf C}$ is 
independent on  any particular choice of coordinate system (e.g., Galactic, Ecliptic, or any 
Euler rotated version of these coordinate systems)  in which the input maps are provided. 
We note, however, the assumption of statistical isotropy is not a necessity in our method. 
If needed, we can also use a covariance matrix compatible to statistically
anisotropic model which may be caused due to non-trivial primordial power spectrum~\citep{Ghosh2016,  Contreras2017}.

\section{Is {\bf C} singular? }
\label{singC} 
As is the case for this work,  rank, $r$, of ${\bf C}$ is less than its dimension $N$. 
The rank of ${\bf C}$ is simply equal to effective number of independent $a_{\ell m}$ 
modes (real and imaginary) that are used in Eqn.~\ref{theory_cov}  to generate each element of the theoretical 
covariance matrix. A quick calculations shows that,  $r = (\ell_{max}+1)(\ell_{max}+2) 
- (\ell_{max}+1) - 4$,   when the summation over multipoles in Eqn~\ref{theory_cov} extends upto 
$\ell = \ell_{max}$. Since we use, $N_{side} = 16$ HEALPix maps in our analysis,  
$\ell_{max} = 2\times N_{side} = 32$ for us, implying $r = 1085$ which is less than dimension of 
${\bf C}$,  which is $N = 3072$. Since ${\bf C}$ is singular we use its generalized inverse 
in Eqn.~\ref{dispersion0}.

\section{Methodology}
\label{method}
\subsection{Input maps and Data Processing}
We use Planck 2015 released LFI 30, 44 and 70 GHz, HFI 100, 143, 217 and 353 GHz frequency maps 
along with the WMAP 9 year difference assembly (DA) maps in our analysis. 
For each of these maps we convert them to spherical harmonic space upto $\ell_{max} = 32$ and 
smooth the resulting $a_{\ell m}$ coefficients by the ratio $B^0_{\ell} P^0_{\ell}/B^i_{\ell} P^i_{\ell}$ 
where $B^i_{\ell}$ and $ P^i_{\ell}$ represent the beam and pixel window functions of the original 
maps whereas  $B^0_{\ell}$ and $ P^0_{\ell}$ represent the corresponding window functions for 
the $N_{side} = 16$ maps.  We take $B^0_{\ell}$ corresponding to  a Gaussian beam of {FWHM = $9^\circ$}. 
We convert the smoothed spherical harmonic coefficients to  $N_{side}=16$ maps using HEALPix supplied 
facility {\tt synfast}. For each of WMAP Q, V and W bands we average all the DA maps for any given 
frequency band.  We convert all these maps  in $\mu K$ (thermodynamic) temperature unit and subtract the corresponding 
mean temperature from each frequency map.  This results in a total of $12$ input maps for foreground removal 
at $N_{side} = 16$.

\begin{figure}
\includegraphics[scale=0.35,angle=90]{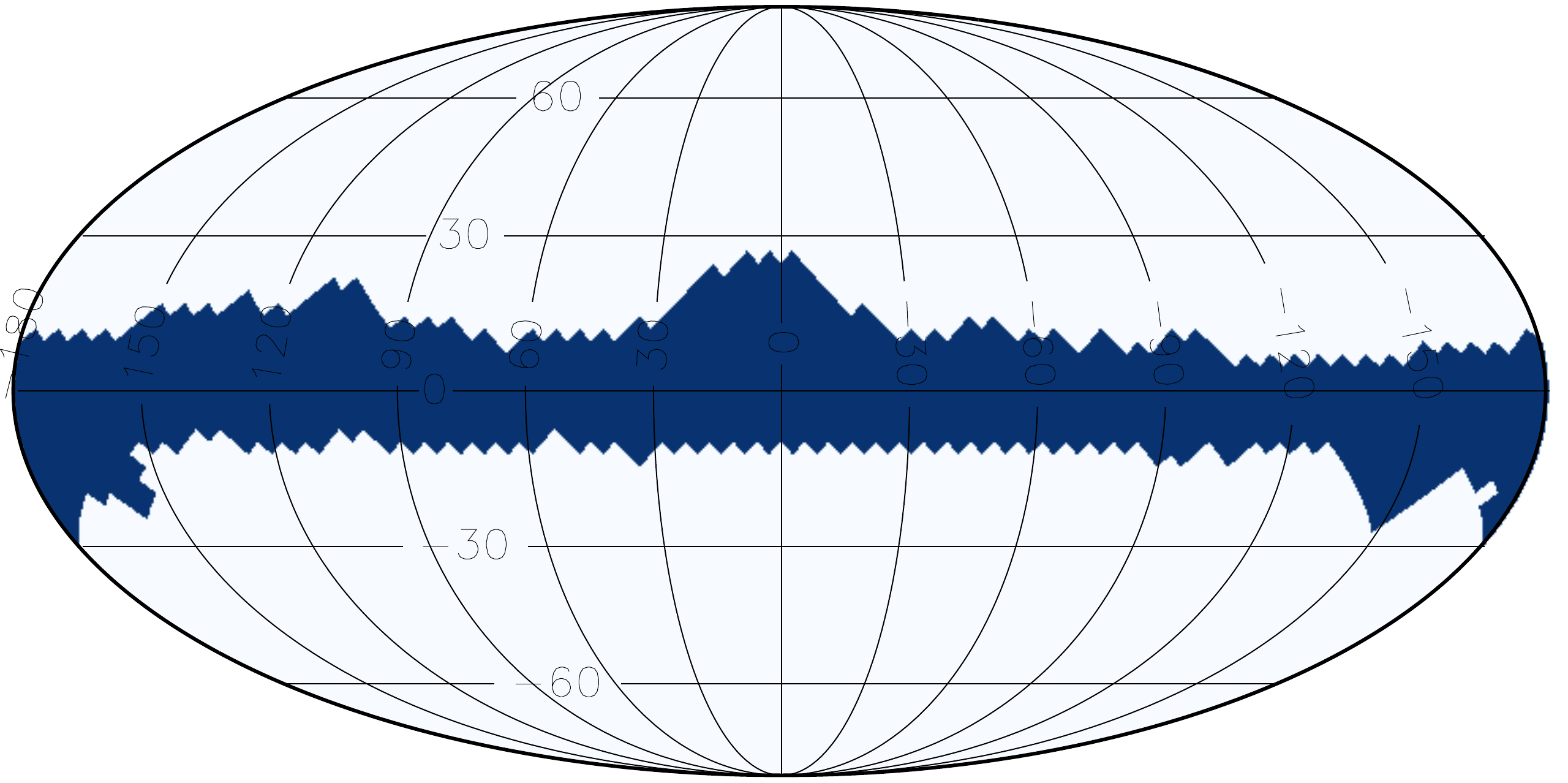}
\caption{The blue region shows sky portions dominated by the strong thermal dust emission and is removed 
by the {\tt ThDust5000}  mask}
\label{mask}
\end{figure}

\subsection{Method-1}
\label{method1}
Since we are interested in a global method  of foreground minimization our aim is to use as much sky 
region as possible to estimate the weights. In the first method we therefore estimate the weights using 
information obtained from the entire sky. We first estimate full sky CMB theoretical covariance matrix using Eqn.~\ref{theory_cov}. 
We obtain ${\bf C}^{\dagger}$ using singular value decomposition of ${\bf C}^{\dagger}$  and applying a cutoff 
of $1.0 \times 10^{-7}$  on the singular values. We show different square blocks across the diagonal of ${\bf C}$  
matrix estimated for the entire sky in Fig.~\ref{covmat}. Non-diagonal elements 
of this matrix show significant coupling between different pixel pairs for a pure CMB map and justifies 
using Eqn.~\ref{dispersion0} for minimization instead of ignoring such correlations as is done in usual pixel 
space ILC approach. Using ${\bf C}^{\dagger}$ we obtain  weights for foreground removal using Eqns.~\ref{AMatrix}
and~\ref{weights}. The cleaned map obtained using these weights is discussed in Section~\ref{Result}.  

\begin{figure*}
\includegraphics[scale=0.68]{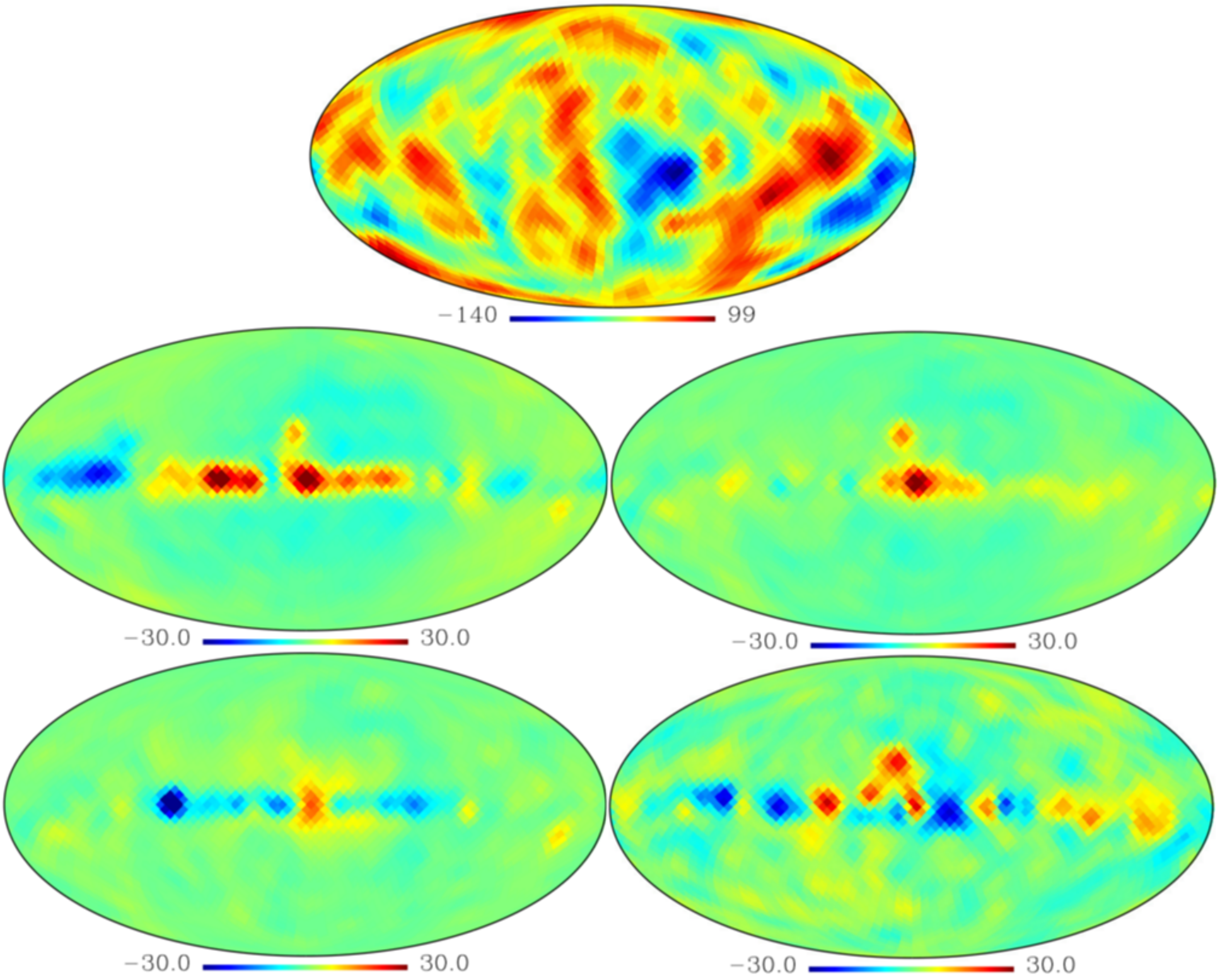}
\caption{Foreground cleaned CMB map ({\tt CMap1}) obtained following Method-1 of this work at $N_{side}=16$ and 
FWHM = $9^{\circ}$ corresponding to Gaussian beam is shown in the top panel. The middle left and middle right 
figures show the difference maps {\tt COMMANDER - CMap1} and {\tt NILC - CMap1} respectively.  The bottom panel 
from left to right show {\tt SMICA - CMap1} and {\tt WMAP ILC - CMap1}. Any residual monopole and dipole 
have been removed from all difference maps to highlight  residuals on cosmologically important scales. 
All color scales are in $\mu K$ thermodynamic temperature unit. }
\label{cmap_fig}
\end{figure*}
 
\subsection{Method-2} 
\label{method2}
Since  when compared with the expected level of CMB temperature anisotropy, the region near galactic plane is 
strongly contaminated by the foregrounds than the outside region, it is desirable to perform foreground removal 
separately on the sky region away from the plane and inside the plane. Moreover the spectral properties of the 
foregrounds vary with sky positions, specifically near the galactic plane. WMAP science team  produce the internal 
linear combination map at $N_{side} = 512$ by dividing the galactic plane into 12 different regions. The sky 
region outside the plane was cleaned in a single iteration. The work of this paper, however, intends to use 
global information from the theoretical CMB covariance matrix and data. Keeping in mind such dual  requirements 
we divide the sky into two regions and clean each as described below. The reason why we divide the sky
into smaller number of regions than an usual ILC approach in pixel space at high resolution,  is that we are 
interested in low resolution maps focusing on the low multipoles. The lack of structures on small scales in the 
input maps ensures that the sky regions need not be too small.

\subsubsection{Sky Division}
To identify the region near the galactic plane that contains strong foreground emissions
we take Planck 353 GHz and 70 GHz frequency maps at $N_{side} = 2048$. We downgrade these
maps to $N_{side} = 256$ and smooth them by the ratio of window functions of a Gaussian 
beam of FWHM = $6^{\circ}$ and the original beam functions of the $N_{side} =2048$ maps 
at the their native resolutions. We subtract resulting reduced resolution 
70 GHz map from 353 GHz map at $N_{side}  = 256$. The difference map contains strong 
emissions from thermal dust at $353$ GHz. We identify pixels  of the difference map
with  values $\ge  5000 $ $\mu K$ and assign a value of unity to them and zero  to 
rest. We downgrade this binary map at $N_{side} = 16$. Finally we reassign all  non-zero  
pixels of the downgraded map a value of zero and the rest to a value of unity. This sky region defined by the zero 
pixel values contains strong thermal dust emissions. The region complementary 
to this strong thermal dust emission is survived after application of  the  {\tt ThDust5000}  mask. The 
sky region removed by  this mask is shown in deep blue color  in Fig.~\ref{mask}.

\subsubsection{Foreground cleaning} 
\label{fg_cleaning}
Based upon the discussions of the previous sections we perform the foreground cleaning following the second 
method in following three steps. 
\begin{enumerate} 
\item{ We estimate the covariance matrix $\tilde {\bf C}$ applicable for the sky region defined 
by the {\tt ThDust5000}  mask. This is done by using Eqn.~\ref{theory_cov} for all the pixel
pairs $(i,j)$ that survive after application of the mask. We estimate $\tilde {\bf C}^{\dagger}$ following the 
same procedure as described in Section~\ref{method1}.}

\item{We use this generalized inverse of the partial sky CMB covariance matrix in Eqn.~\ref{AMatrix} to obtain elements of the partial 
sky $\tilde{\bf A}$ matrix. Using this partial sky matrix in Eqn.~\ref{weights} we obtain 
the weights corresponding to the {\tt ThDust5000} sky region. Using these weights we obtain
the cleaned {\tt ThDust5000} sky region.  }

\item{Now we replace the {\tt ThDust5000} sky region of all foreground contaminated input maps 
by the cleaned region obtained above. The resulting 12 maps have their galactic regions yet 
to be cleaned and strongly contaminated by the foregrounds. To clean the galactic region  we 
repeat steps 1 and 2 above over the full sky.  The cleaned map obtained at this point is the full-sky
cleaned map obtained by Method-2.}  
\end{enumerate}

\begin{figure}
\includegraphics[scale=0.35,angle=90]{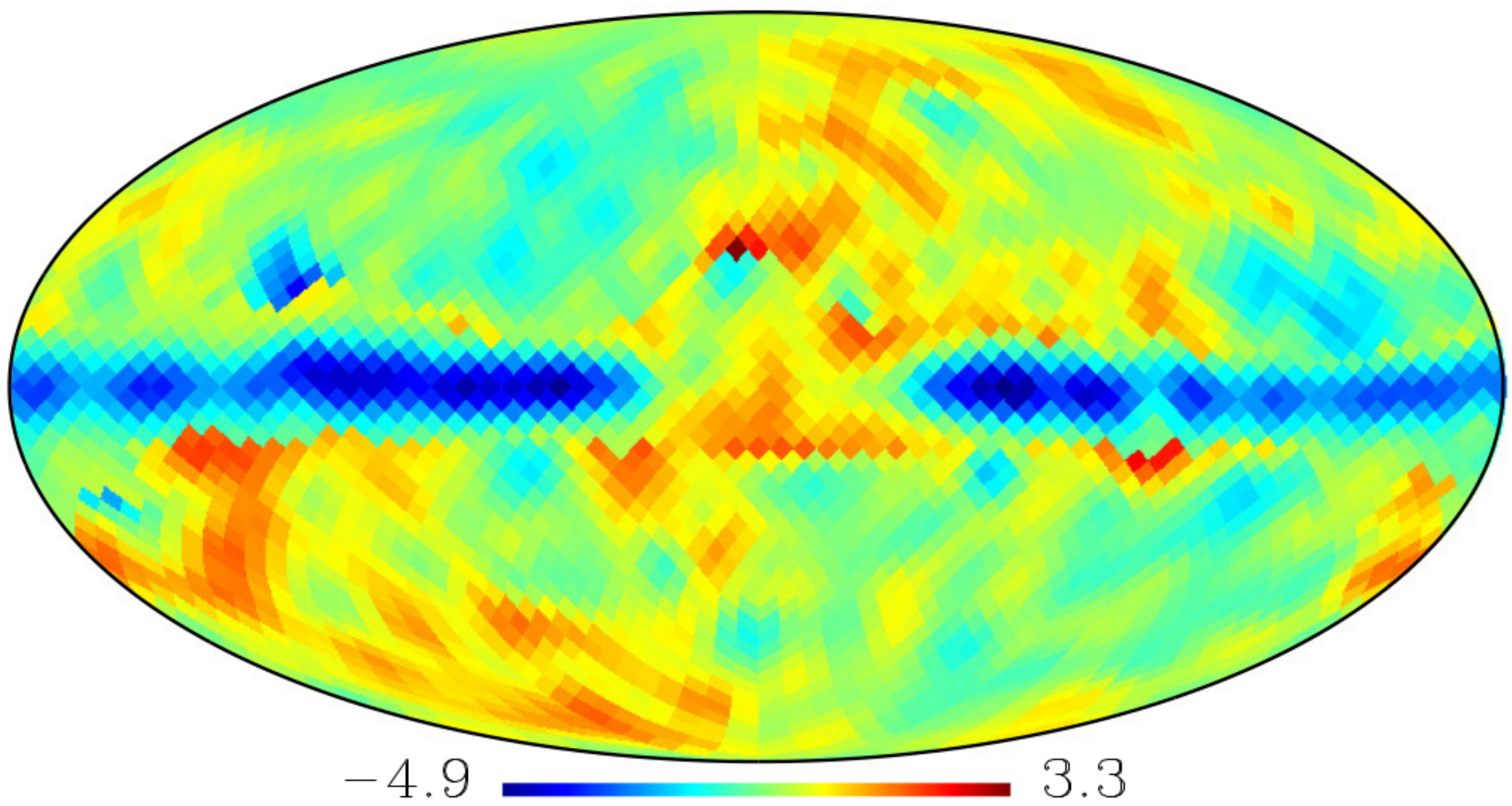}
\caption{Difference of full sky cleaned CMB maps ({\tt CMap2-CMap1}) obtained from Method 1 and Method 2. {\tt CMap2} appears 
to have lesser foreground contamination along  the both sides of the galactic plane. Temperature scale is in $\mu K$ 
thermodynamic unit.}
\label{diffcmap}
\end{figure}
 
\section{Results} 
\label{Result}
\subsection{Cleaned Maps} 
Using the first method the weights for different WMAP and Planck channels become 
$-0.093$, $0.226$, $0.424$, $-0.392$, $-0.859$, $-0.105$, $0.195$, 
$0.390$, $0.890$, $0.906$, $-0.607$, $0.0245$ in the increasing order 
of frequency of the 12 input maps from  23 to 353 GHz. We use these weights  to  
linearly combine the 12 input maps to estimate the cleaned CMB map at $N_{side} =16$ 
and at Gaussian beam resolution of FWHM = $9^\circ$ (henceforth we call this cleaned 
map {\tt CMap1}). We show the {\tt CMap1} in the top  panel of Fig.~\ref{cmap_fig}. 
Visually the {\tt CMap1}  does not contain any foreground residuals.  
We compare this map  with other foreground minimized CMB maps each of which is obtained
by employing a different algorithm at higher beam and pixel resolutions, as reported in the literature.  
{\tt COMMANDER} CMB map was obtained following  joint estimation of CMB and all foreground components, {\tt NILC} 
CMB maps was obtained by employing an internal linear combination algorithm in the  needlet space and 
{\tt SMICA} CMB map was obtained by using spectral matching technique (e.g., see~\cite{Planck2016_CMB} for 
detailed discussion about these maps). WMAP science team produced  a CMB map  by using usual ILC approach 
in pixel space~\citep{Hinshaw_07,Gold2011}. We  downgrade these high resolution maps at $N_{side} = 16$ 
and bring them to a common beam resolution of $9^{\circ}$. We show the difference of {\tt CMap1} from resulting 
{\tt COMMANDER} and {\tt NILC} maps respectively in the middle left and right panel of Fig.~\ref{cmap_fig}. The lower 
left and right panel show differences of {\tt CMap1} from  {\tt SMICA}  and WMAP {\tt ILC} maps. Since monopole and dipoles 
are not of any cosmological interests we have removed any residual dipole and monopole from all the 
four difference maps shown in this figure.  Clearly our cleaned 
CMB map matches well with these cleaned CMB maps in the higher galactic plane.  Along the galactic plane 
we find some differences. However, as one can easily make out such  difference along the galactic plane exists for 
any pair of all five low resolution CMB maps discussed in this section.   

\begin{figure}
\includegraphics[scale=0.9]{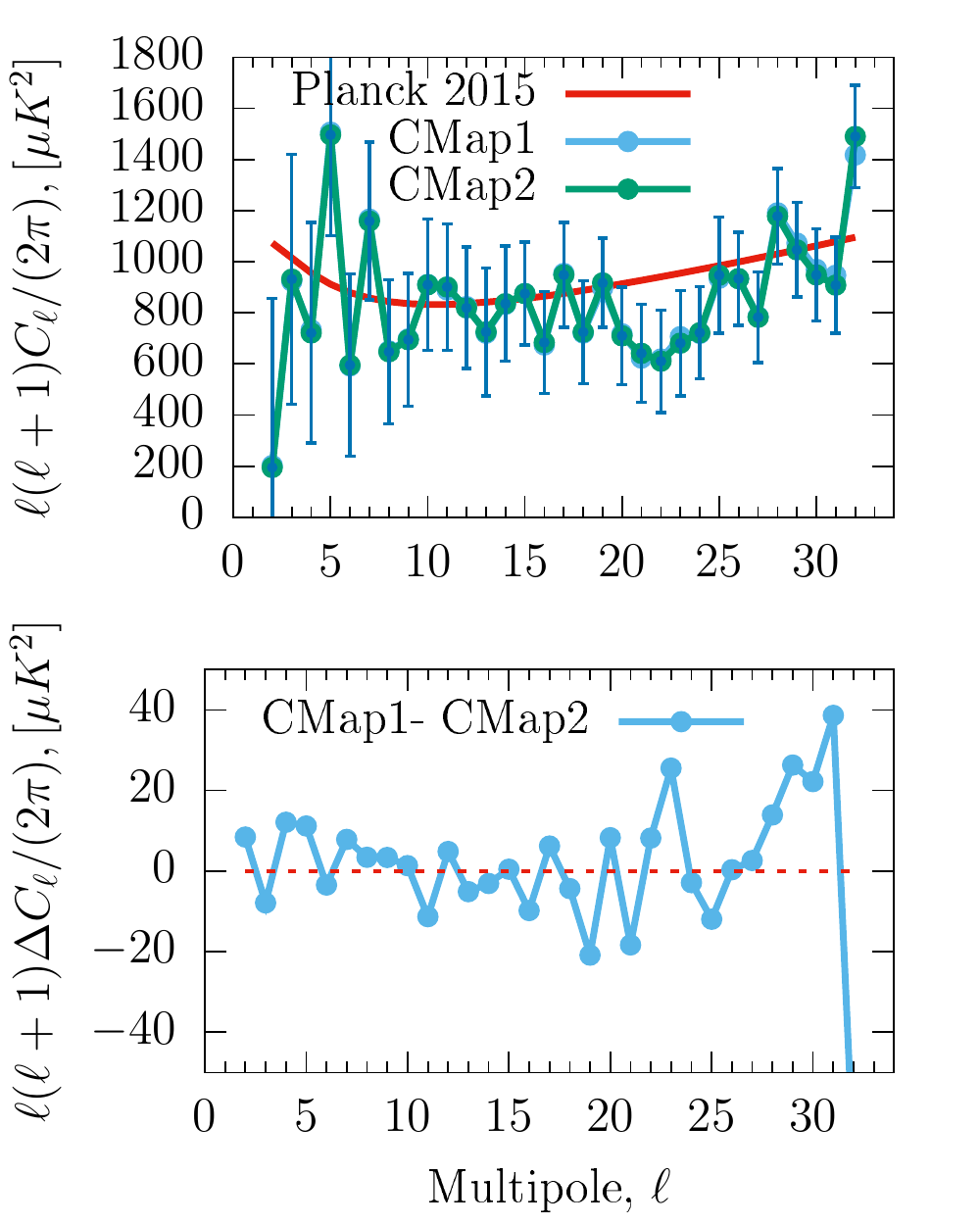}
\caption{Top panel shows  estimates of CMB angular power spectrum obtained from  full sky region
of {\tt CMap1} (Method-1), {\tt CMap2} (Method-2)  along with the Planck 2015 theoretical LCDM power spectrum. Both these 
observed spectra of Method-1 and Method-2  match well with each other. The error-bars are compatible to Method-2. The bottom panel shows 
difference of spectra obtained from these two methods. The dashed line shows the zero level of the power 
spectrum. }
\label{pow_spec_fig}
\end{figure}

\begin{figure*}
\includegraphics[scale=0.68]{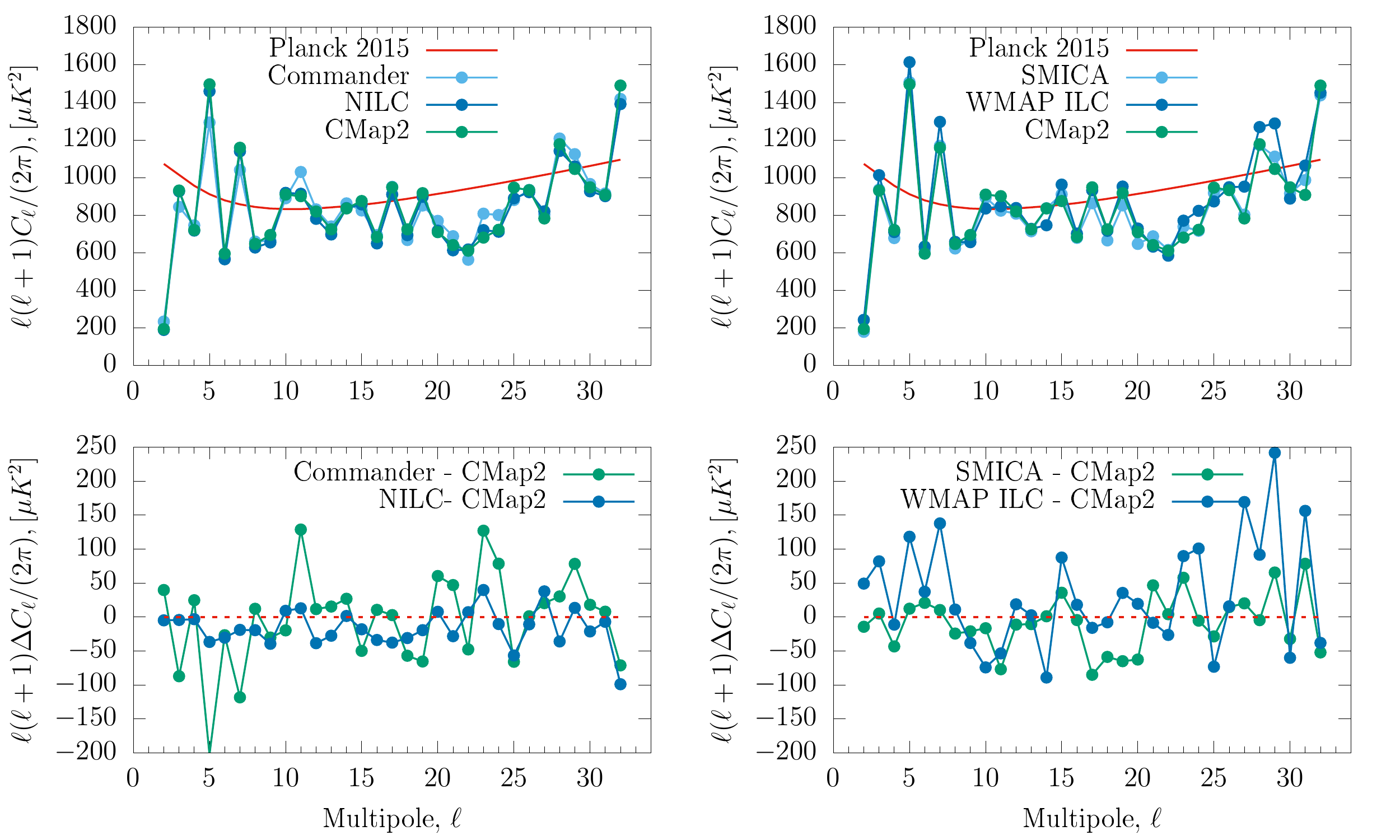}
\caption{Both figures of top panel shows  CMB angular power spectrum obtained from {\tt CMap2} of this work with the 
same estimated from  other foreground cleaned CMB maps as mentioned. The bottom panel shows the difference 
of {\tt CMap2} spectrum with those obtained from the other foreground cleaned maps of the top panel. The zero level is shown 
by the red-dashed line.}
\label{pow_spec_fig1}
\end{figure*}

Following the second method we recover a cleaned map ({\tt CMap2}) similar to {\tt CMap1}. The weights for the sky region 
survived after application of {\tt ThDust5000} mask are  $-0.066$,    $0.083$,    $0.500$,    $-0.306$,   $-0.562$,
$ -0.757$,   $0.021$,  $0.917$,  $0.876$,   $0.948$,   $-0.684$ and  $0.031$ respectively for different frequencies 
increasing from 23 to 353GHz (e.g., see step 2 of Section~\ref{fg_cleaning}). The corresponding weights for the full 
sky (step 3 of Method-2) linear combination  are  $-0.084$,  $0.240$,  $0.414$,  $-0.399$,  $-0.994$,  $0.100$, $0.217$,
$0.277$,  $0.913$, $0.896$,  $-0.604$ and   $0.024$ respectively. A common feature of the weights for both these 
regions is that strongly contaminated frequency maps (e.g., K1 band or 353 GHz) get  low (negative or positive)
weights to cancel out foregrounds from all frequencies. The {\tt CMap2}  matches closely 
with the {\tt CMap1}. We show the difference  {\tt CMap2 - CMap1}   in 
Fig.~\ref{diffcmap}. Clearly the Method-2 has slightly less foreground residuals along the both sides of the galactic 
plane at the expense of some additional detector  noise residuals along the ecliptic plane. We compare the 
full sky power spectra of {\tt CMap1} and {\tt CMap2} along with other CMB spectra in Section~\ref{pow_spec}.

%\begin{figure}
%\includegraphics[scale=0.9]{cl2}
%\caption{Top panel shows  estimates of CMB angular power spectrum obtained from  full sky region
%of {\tt CMap1} (Method-1), {\tt CMap2} (Method-2)  along with the Planck 2015 theoretical LCDM power spectrum. Both these 
%observed spectra of Method-1 and Method-2  match well with each other. The error-bars are compatible to Method-2. The bottom panel shows 
%difference of spectra obtained from these two methods. The dashed line shows the zero level of the power 
%spectrum. }
%\label{pow_spec_fig}
%\end{figure}

\subsection{Power Spectrum}   
\label{pow_spec}
We show the CMB angular power spectra after corrections of beam and pixel effects obtained from full sky of 
{\tt CMap1} and {\tt CMap2} in the top panel of Fig.~\ref{pow_spec_fig}. The theoretical CMB angular
power spectrum is shown in red line to guide the eye. The error-bars show the reconstruction error in power
spectrum obtained from Method-2 and agree well with the cosmic variance induced errors (e.g., see Section~\ref{validity}). The bottom panel of this figure 
show difference of the spectra of these two maps. As we see from this figure both spectra match very well with 
each other. Such close match is also expected from the very small difference between the two cleaned maps 
as shown in Fig.~\ref{diffcmap}. This results suggest that our new ILC approach is very weakly dependent on the 
sky divisions. This justifies following a global approach of foreground cleaning on large angular scales on the 
sky, as is done in this work.  However, since method 2 simultaneously follows a  global approach and performs 
foreground removal in an iterative fashion, we treat CMB angular power spectrum of {\tt CMap2} as the 
main power spectrum of this work estimated using  low resolution Planck and WMAP maps.

We compare full sky CMB angular power spectrum obtained from {\tt CMap2} with the corresponding spectra 
obtained from {\tt COMMANDER, NILC, SMICA} and WMAP {\tt ILC} maps. We show these spectra in top left and 
top right panels of Fig.~\ref{pow_spec_fig1}. Also shown in these two panels is CMB theoretical angular power 
spectrum obtained from Planck 2015 results. The bottom panels of this figure show the difference of angular power 
spectra of this work with the other spectra of the corresponding top panels. As we see from this figure 
the CMB angular power spectrum from {\tt CMap2} match closely with the angular spectra  of these cleaned maps. 
A similar result was obtained considering CMB angular power spectrum from {\tt CMap1} also. It is noteworthy 
 that power spectra of {\tt CMap2}  and  {\tt NILC} map agree excellently  for the entire multipole range 
$2 \le \ell \le 32$.

\section{Monte Carlo Simulations}
\label{validity}
\subsection{Input CMB, Foreground and Noise Maps} 
\label{SimInput}
We validate the methodology for the first and second methods by performing detailed Monte Carlo simulations of the entire foreground removal 
and power spectrum estimation procedures. For this purpose we first generate foreground maps at different 
WMAP and Planck frequency bands of this work.  The free free, synchrotron and thermal dust emissions at 
different frequencies are first obtained at $N_{side} = 256$ and beam resolution $1^\circ$ following the procedure 
as described in~\cite{Sudevan2017}\footnote{Unlike the work of~\cite{Sudevan2017} in the current work we 
use a spatially constant spectral index ($\beta_s = -3.00$) for synchrotron component for all WMAP and Planck frequencies.}. 
We then downgrade the pixel resolution of each component map to $N_{side} = 16$ 
and smooth each one by Gaussian beam function of FWHM = $\sqrt{540^2 - 60^2} = 536.66^{\prime}$ so as to 
bring all component maps for all frequency maps to the common resolution of $9^\circ$. We generate CMB temperature 
anisotropy maps at $N_{side} = 16$ and FWHM = $9^\circ$ by using the theoretical CMB power spectrum consistent 
with cosmological parameters obtained by~\cite{PlanckCosmoParam2016}. The procedure to generate the detector 
noise maps remains similar to~\cite{Sudevan2017}. Following the same procedure as described by these authors, we first 
generate noise maps  at $N_{side} = 512$ (for WMAP DA maps)  or $1024$ and $2048$ (for Planck frequency maps).
 We then convert these maps to spherical harmonic space upto $\ell_{max} = 32$, 
and multiply the resulting spherical harmonic coefficients by the ratio of the window function corresponding to 
FWHM = $9^\circ$ and the native beam window function of each WMAP DA (or Planck frequency bands). For WMAP 
Q, V and W band each,  we average the DA noise maps to generate a single noise map corresponding to the 
given frequency band. We generate a set of $200$  noise maps for each of $12$ frequency maps of 
our analysis. Each of these noise maps have uncorrelated noise properties. We add the CMB, foreground and noise
maps generated above to obtain a set of frequency maps that represent realistic observations of WMAP and Planck 
missions at $N_{side} = 16$ and FWHM = $9^\circ$. We generate a total of  $200$ such sets of input frequency maps for Monte Carlo 
simulations.  

\begin{figure}
\includegraphics[scale=0.65]{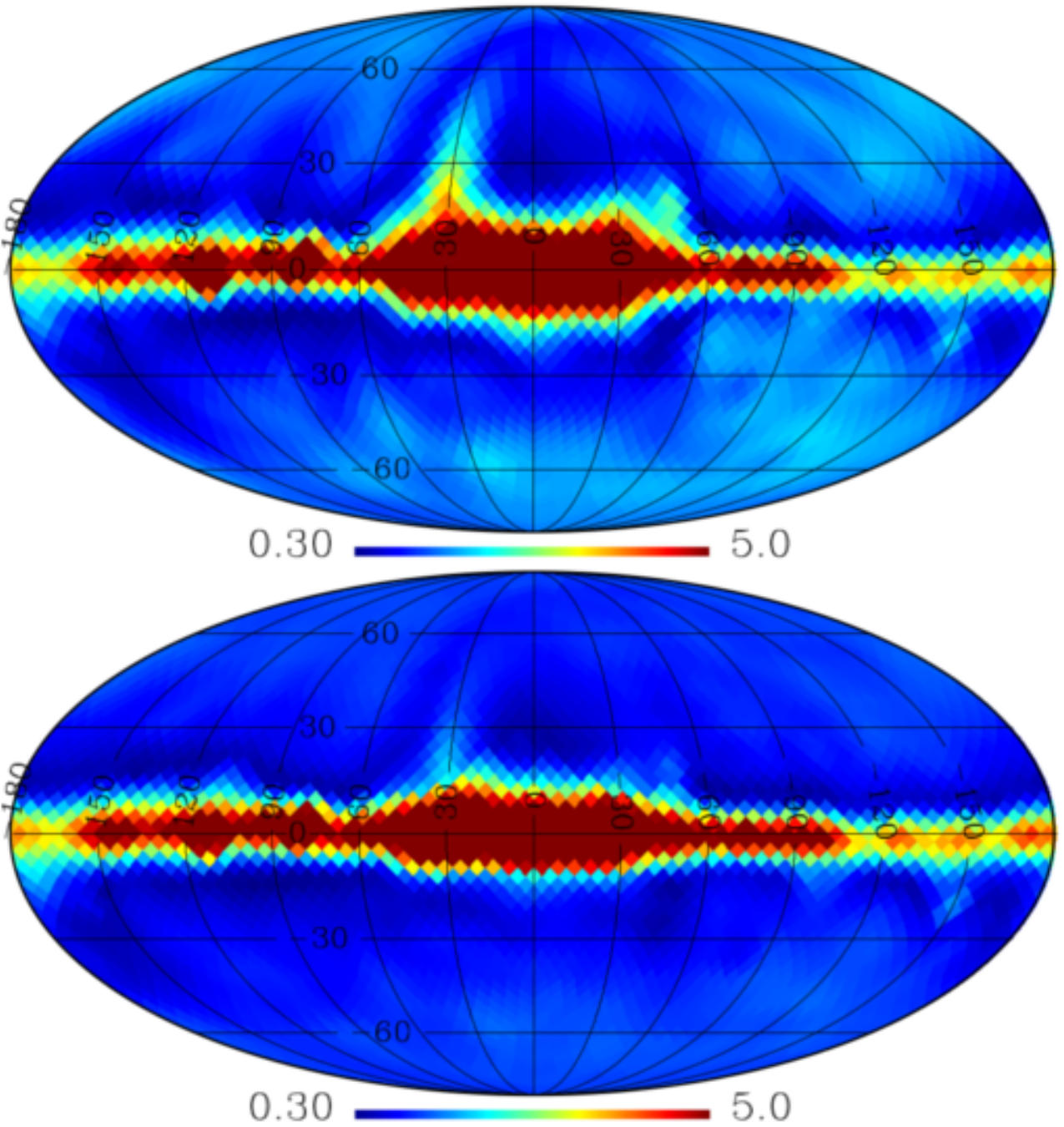}
\caption{Top panel shows standard deviation map obtained from the difference of foreground minimized CMB map and corresponding 
randomly generated input CMB map using $200$ Monte Carlo simulations of foreground minimization following Method-1 as described  
in Section~\ref{method1}. The bottom panel shows the standard deviation map obtained for the $200$ Monte Carlo 
simulations of Method-2 (e.g., see Section~\ref{method2}). All units are in $\mu K$ thermodynamic temperature. The reduction in reconstruction error for these 
two methods is discussed in Section~\ref{sim_results}.} 
\label{err_map}
\end{figure}

\subsection{Results}
\label{sim_results}
\subsubsection{Reconstruction Error in Cleaned Maps}
\label{CMBErr}
If the input CMB map for the $i$th Monte Carlo simulations is denoted by $T_i(p)$, where $p$ denotes the pixel index, 
and the corresponding foreground minimized CMB map is $T^{\prime}_i(p)$, the map representing reconstruction error 
for the particular simulation is then given by $\Delta T_i(p) = T^{\prime}_i(p) - T_i(p)$. We estimate the standard 
deviation map using all $200$ error maps for each of our two methods of this work. The error-maps for method 1 and 
2 are shown respectively is top and bottom panels of Fig.~\ref{err_map}. As seen from this figure, using the iterative
method reduces the reconstruction error in the north and southern hemisphere  towards the galactic center region.   
Also seen from this figure is  lower reconstruction error near  the  north polar spur region. The average variance per pixel 
over full sky for method 1 (estimated from the top panel  Fig.~\ref{err_map}) is  $6.41 \mu K^2$ compared to a value 
of $5.25 \mu K^2$ for the method 2 (bottom panel). Corresponding average variances for {\tt ThDust5000} mask region 
are  $1.75$ and $0.97 \mu K ^2$ respectively. For galactic region not covered by the thermal dust mask
the average variances become $22.84$ and $20.36 \mu K^2$ respectively for the Method-1 and Method-2. We conclude 
both methods work with comparable efficiencies, however, the second method performs better than the first 
method in terms of foreground removal.

\subsubsection{CMB Angular Power Spectrum} 
\begin{figure}
\includegraphics[scale=0.85]{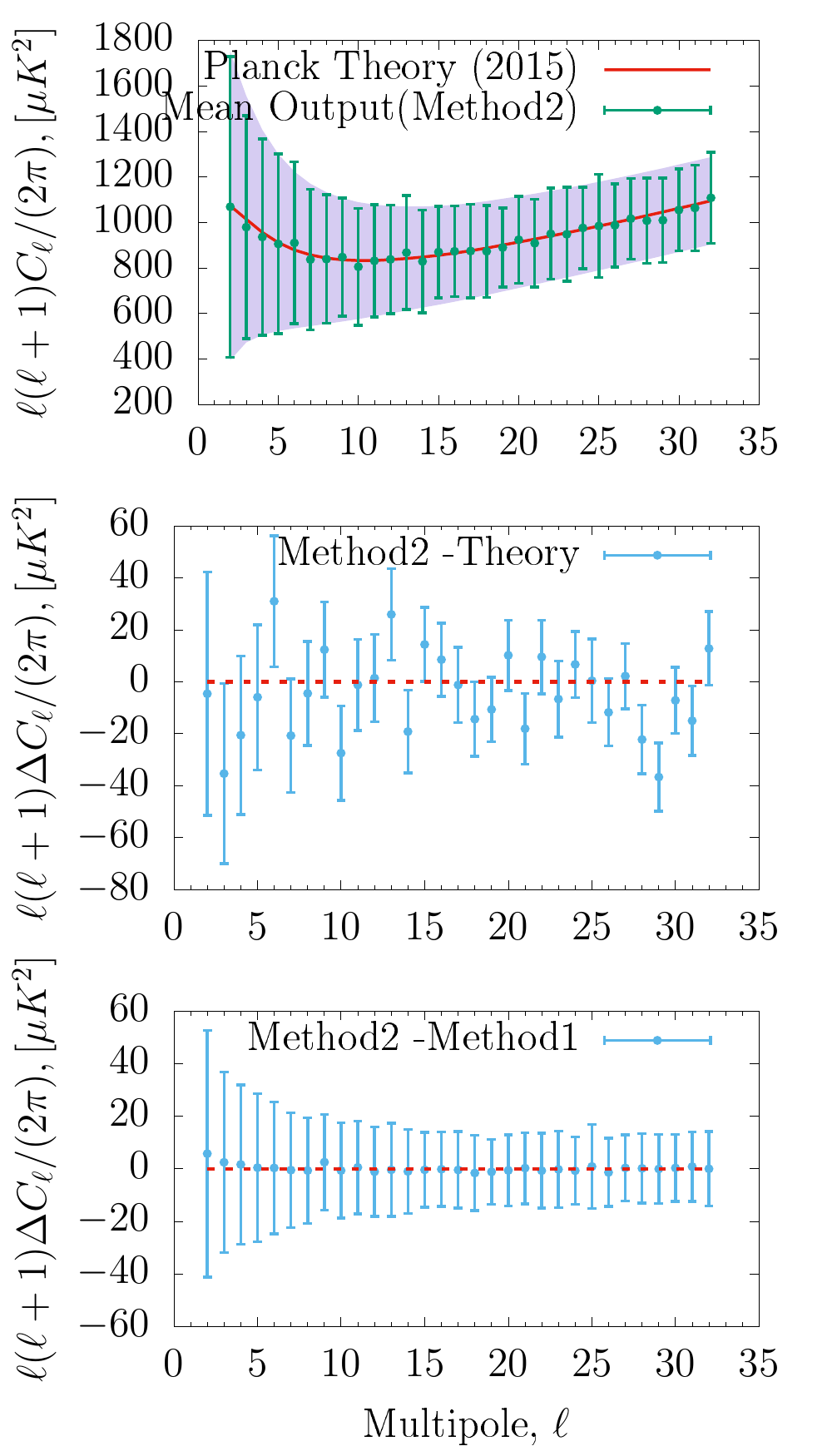}
\caption{Top panel shows the mean (in green) of $200$ full sky CMB angular power spectra obtained from Monte Carlo simulations 
of Method-2 of this work along with the theoretical CMB power spectrum (red line). The error bar computed from cosmic variance estimated 
from the theoretical power spectrum shown by the filled region. The reconstruction error on  cleaned CMB power spectrum obtained from
any one of the simulations is shown in green. The middle panel shows a close comparison of mean CMB angular power spectrum following 
Method-2 and theoretical CMB power spectrum. The error-bars of this plot represents error on the foreground cleaned mean CMB spectrum.
The bottom panel shows the difference between the mean spectra of Method-2 and Method-1 along with the error bars applicable for 
mean spectrum of Method-1.  }  
\label{sim_pow_spec}
\end{figure}  

Using $200$ foreground cleaned maps obtained from Monte Carlo  simulations of foreground removal and subsequent CMB 
angular power spectrum estimation over the complete sky region we assess reconstruction error in cleaned CMB power 
spectra obtain using Method-1 and Method-2. In top panel of Fig.~\ref{sim_pow_spec} we plot mean CMB angular power
spectrum  (green points) obtained following Method-2 along with the standard deviation of the cleaned power spectrum 
for any one of the simulations.  
The mean foreground cleaned power spectrum agrees well with the theoretical CMB power spectrum (red line) which is used to
generate random (and isotropic) CMB realizations. The cosmic variance  error limit is shown by the colored band around 
the theoretical CMB power spectrum. The close match of cosmic variance and the reconstruction error on the cleaned 
power spectrum at each multipole implies that the recovered angular power spectrum is only cosmic variance limited and reconstruction 
error due to foreground residuals (plus any error induced by detector noise) is a sub-dominant source of contamination 
on the angular scale chosen in this work. In the middle panel of Fig.~\ref{sim_pow_spec} we closely investigate any 
reconstruction biases that may exist in the foreground cleaned power spectrum of Method-2 by plotting the 
difference between foreground minimized mean CMB power spectrum  and the CMB theoretical power spectrum. The error
bar at each multipole plotted in this panel is applicable for the mean CMB angular power spectrum and therefore they are
obtained by scaling the corresponding reconstruction error of top panel  by $1/\sqrt{N_{sim}}$ where the number 
of simulations, $N_{sim} = 200$. For all the multipoles except ($\ell = 29$) the significance of any difference 
between the mean cleaned spectrum and the theoretical CMB spectrum is less than $2\sigma$. For $\ell=29$ the 
significance  of deviation is  $2.8\sigma$. This shows that power spectrum obtained from Method-2 has no
significant bias that may arise due to imperfect foreground residuals. The bottom panel of Fig.~\ref{sim_pow_spec} shows the 
difference between mean CMB power spectra obtained from Method-1 and Method-2. The error-bars of this plot is 
computed from foreground cleaned maps of Method-1 and they are applicable for the mean power spectrum. Clearly
mean spectra obtained by the two methods of this work agree very well with each other. Both methods produce
comparable error-bars as well.    

\section{Advantage of global ILC method at low resolution}
\label{advantage}
The global ILC method has two very important advantages over the usual ILC method in pixel space that does not take into account
prior information about CMB theoretical covariance matrix. First, the globally cleaned CMB map has less reconstruction 
error at each pixel. Second, the usual ILC approach (without using the covariance information) at low resolution
leads to a bias in the power spectrum which remains absent in the proposed methods of this work. The cause of these
limitations in usual ILC approach at low resolution analysis is a chance-correlation between the CMB and foreground (and 
detector noise) components which can not be ignored over large scales of the sky. In this section we discuss about the
advantages of our approach.  

\begin{figure}
\includegraphics[scale=0.35, angle=90]{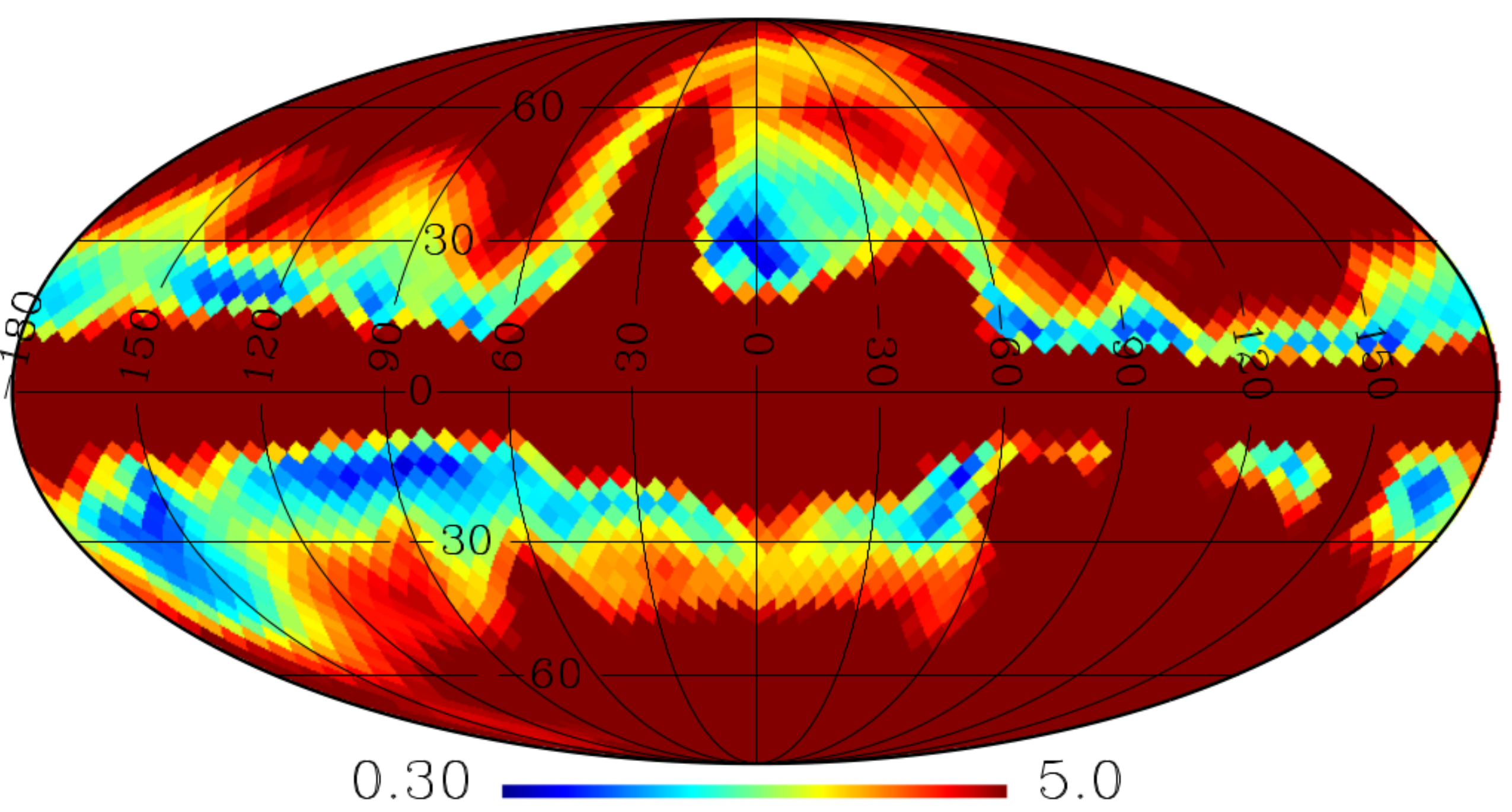}
\caption{The standard deviation map indicating the reconstruction error for usual pixel space ILC approach over the entire 
sky at $N_{side} = 16$ and $9^{\circ}$ resolution. Large reconstruction error compared to the methods (e.g., see Fig.~\ref{err_map}) 
of this work is seen. Unit is in $\mu K$ thermodynamic.}
\label{ErrOldILC}
\end{figure}

Using the simulated frequency maps  at $N_{side}= 16$ and $9^{\circ}$ resolution (e.g., Section~\ref{SimInput}) we perform 
$200$ Monte Carlo simulations over the complete sky region using usual ILC approach, wherein no CMB covariance matrix is used.
The error map in CMB reconstruction is then computed in the same fashion as discussed in Section~\ref{CMBErr}.  The standard
deviation map is plotted in Fig.~\ref{ErrOldILC} which indicates a strong residual, not only on the the galactic plane, but
also in higher galactic latitudes. Unlike the small variance per pixel reported in Section~\ref{CMBErr} the average 
variance per pixel for Fig.~\ref{ErrOldILC} is large ($89.17 \mu K^2$). This clearly demonstrates the first advantage, 
i.e., sharp decrease  in reconstruction error of cleaned CMB map, when we incorporate prior information about theoretical covariance of CMB component.  

\begin{figure}
\includegraphics[scale=0.9]{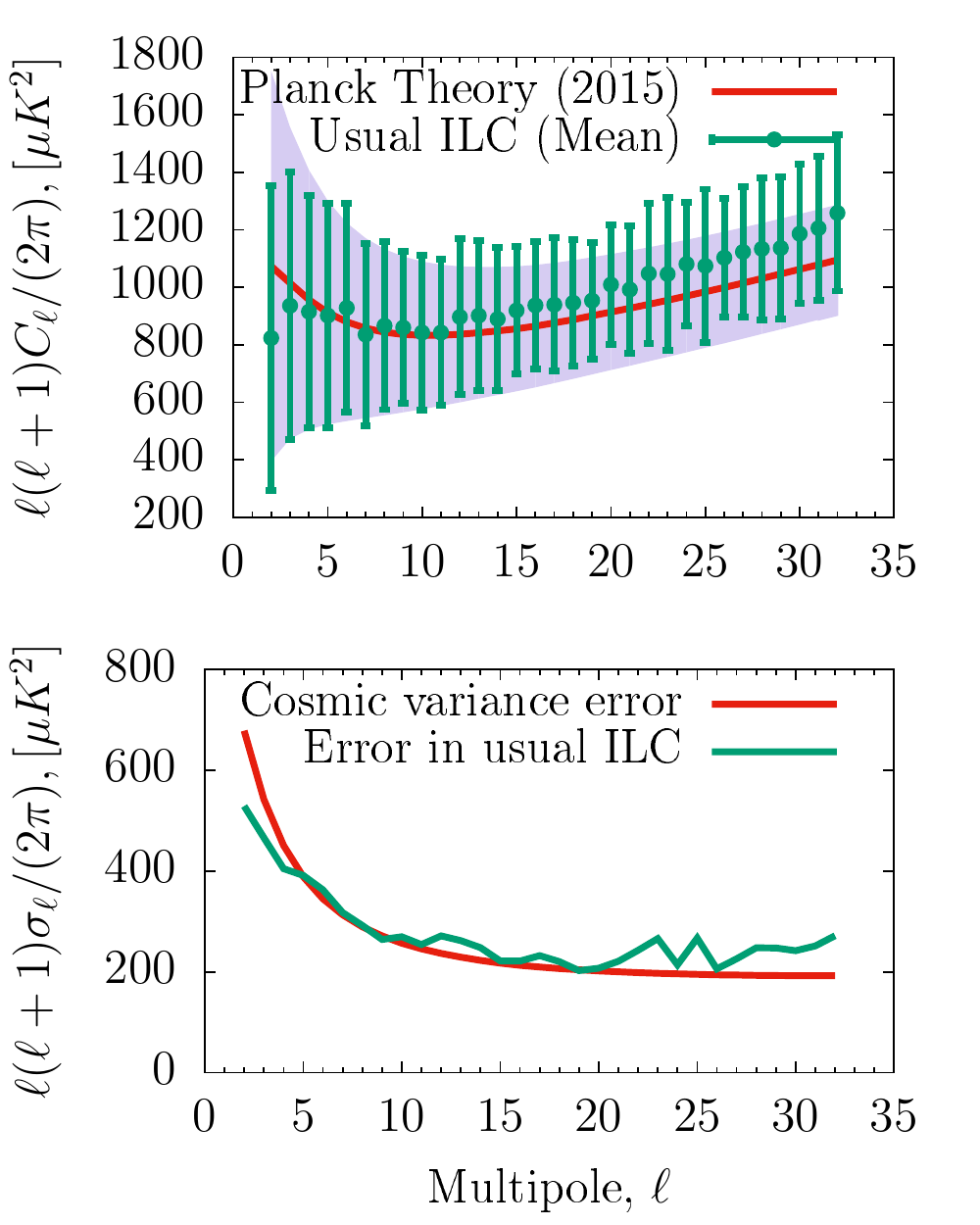}
\caption{Top panel shows the mean CMB power spectrum (in green) obtained from $200$ Monte Carlo simulations of usual ILC 
approach over the entire sky on low resolution maps as discussed in Section~\ref{advantage} along with the theoretical 
CMB angular power spectrum (red line) consistent with Planck 2015 results. The filled color band shows the cosmic variance 
excursion limit of the observed CMB angular power spectrum. The green error-bars show reconstruction  error in the 
cleaned power spectrum at different multipoles.  The bottom panel closely compares the reconstruction error-bars 
with the cosmic variance induced errors.  Residuals in the cleaned maps cause larger than cosmic variance error 
starting from $\ell \sim 10$. For low multipoles, $\ell \le 4$ reconstruction error becomes less than cosmic variance 
induced error since the cleaned power spectrum at low  multipoles is biased low due to a chance correlation of CMB 
with foregrounds (and detector noise).  }
\label{ErrOldILC1}
\end{figure}

The larger reconstruction error in cleaned maps in usual ILC approach, causes a significant bias in the power spectrum which 
is a quadratic function of the data. We show the mean power spectrum computed from $200$ Monte Carlo simulations of usual 
ILC approach over the entire sky in green in top panel of Fig.~\ref{ErrOldILC1} along  with the Planck 2015 theoretical 
power spectrum which is used to generate the input CMB maps. Clearly a positive bias exist due to imperfect foreground 
residuals  in the cleaned spectrum starting from multipole $\ell = 8$. Another interesting feature of the top panel is 
existence of a negative bias for $\ell \le 5$. Such negative bias is expected and was first reported by~\cite{Saha2006} and 
is discussed extensively in~\cite{Saha2008} (see also~\cite{Sudevan2017} for such bias in high resolution analysis) for 
multipole space ILC methods. In fact, observing the error pattern of Fig.~\ref{ErrOldILC} it is likely that a positive bias 
due to residual foregrounds exists even at low multipoles, $\ell \le 8$ on the top of the additional negative bias in this 
multipole range. The bottom panel of Fig.~\ref{ErrOldILC1} compares the reconstruction error in the cleaned power spectrum
with the error due to cosmic variance alone. Starting from multipole $\ell \sim 10$ we see that the error in usual ILC 
power spectrum becomes larger than the cosmic variance induced error. Interestingly, due to existence of negative bias 
at the low multipoles the error in cleaned spectrum become biased low for $\ell \le 4$. The bias existing in the cleaned 
power spectrum of the usual ILC approach at low resolution along with larger error in reconstructed power spectrum from 
this approach justifies our second point of advantage (discussed at the beginning of the current section) of the new approach 
described in this article.

\begin{figure}
\includegraphics[scale=0.35, angle=90]{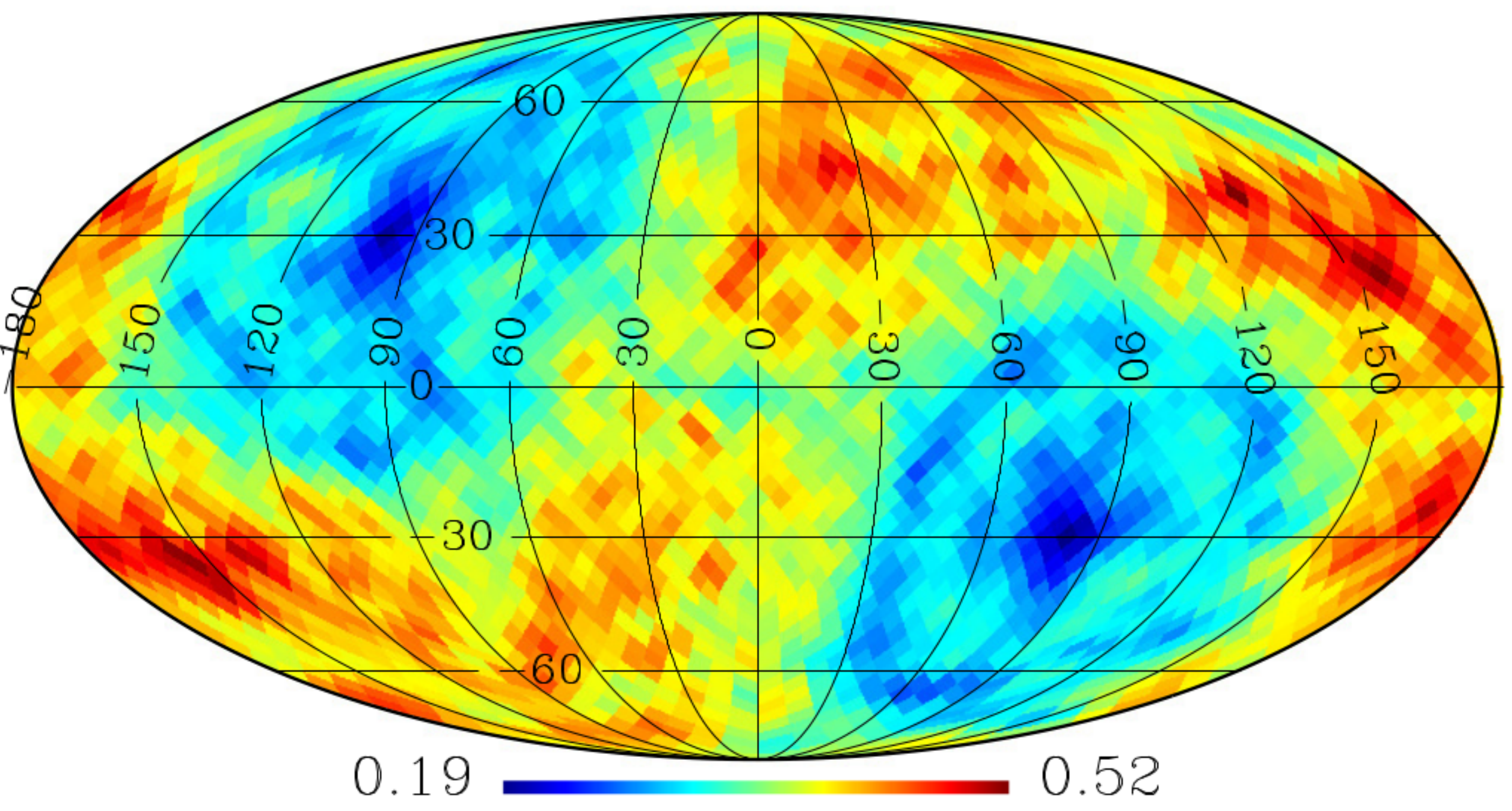}
\caption{The standard deviation map computed from the difference  of cleaned CMB maps and  corresponding 
input CMB maps  using ILC method, when the weights are obtained from Eqn.~\ref{weights2}. The reconstruction 
error follows a noise pattern and is much smaller compared to Fig.~\ref{ErrOldILC} when CMB-foreground 
chance correlation affects the weight estimation. Unit is in $\mu K$ thermodynamic.}
\label{ErrCf}
\end{figure} 

\section{Role of CMB-Foreground (or CMB-Noise) Chance Correlation}
\label{chancecorr}
\begin{figure*}
\includegraphics[scale=0.6]{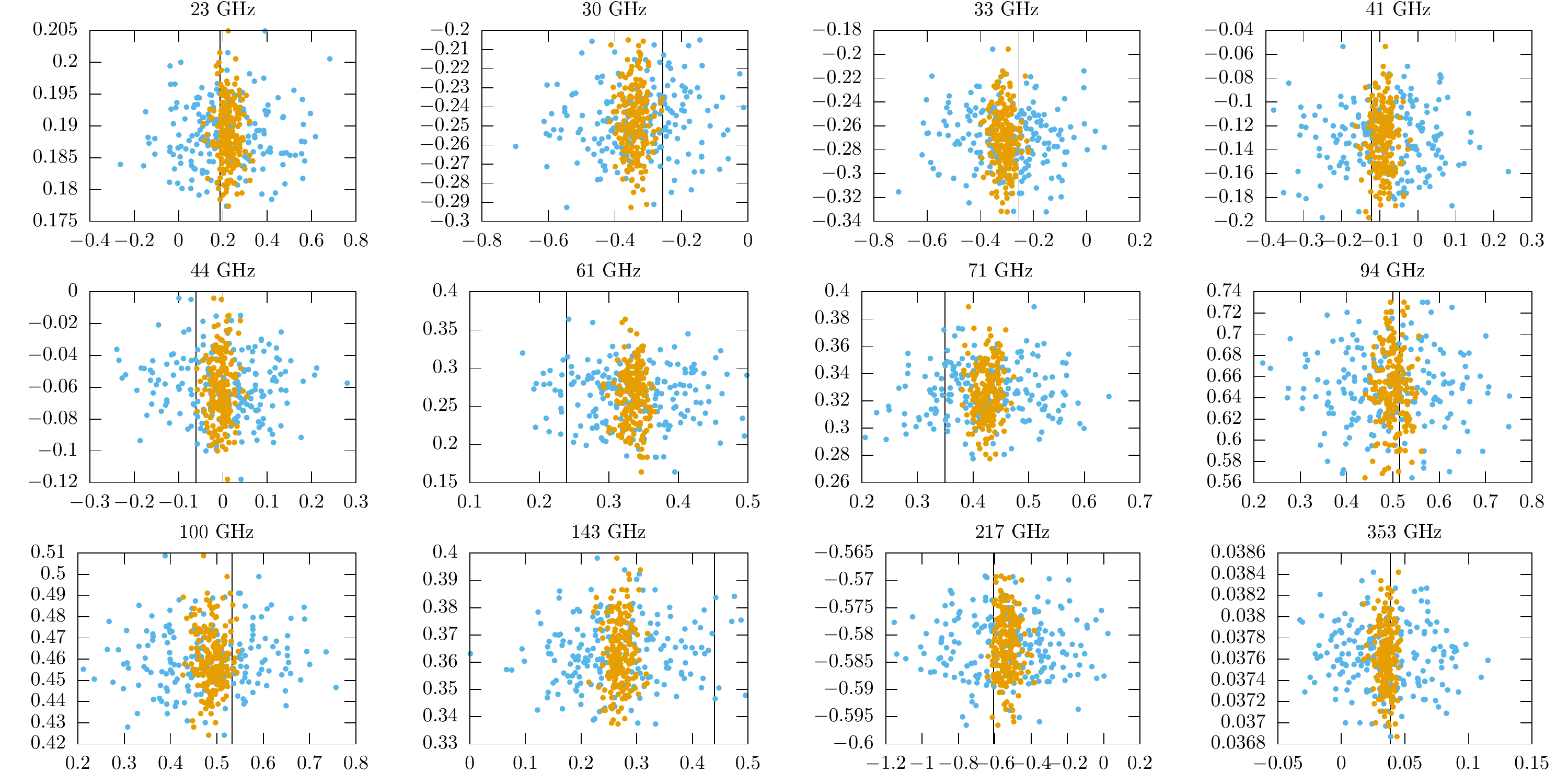}
\caption{Scatter plot obtained from Monte Carlo simulations showing lower dispersion (along horizontal axis) of  
weights (yellow) when we follow global ILC method with prior information from theoretical CMB covariance matrix 
on large scales on the sky for different WMAP and Planck frequency bands. The y coordinates of blue or yellow points 
represent weights obtained from Monte Carlo simulations using Eqn.~\ref{weights2}. The larger dispersion of blue points along the horizontal
axes causes larger reconstruction error in cleaned maps for usual ILC methods at low resolution. The vertical 
lines represent values of weights obtained using Eqn.~\ref{weights2} but without any detector noise in the simulations. }
\label{W}
\end{figure*}

Having discussed in the previous section the advantages of the global ILC method of this work we now 
focus on the cause of excess residuals in the usual ILC method when applied to low resolution maps.   
If we apply usual ILC method on the input maps described in Section~\ref{formalism} variance of the 
cleaned map becomes, 
\begin{eqnarray}
\hat \sigma^2 = {\bf W }\hat{\bf  C} {\bf W}^T\,, 
\end{eqnarray}   
where $\hat {\bf C}$ is an $n \times n$ matrix representing the covariance between different input frequency 
maps (from which mean temperature anisotropies corresponding to each frequency is already subtracted).  
Similar to Eqn.~\ref{weights} the set of weights that minimizes variance of the cleaned map subject to the constraint CMB is preserved 
is given by~\citep{Saha2008, Tegmark2003, Tegmark96}, 
\begin{eqnarray}
{\bf W} = \frac{ {\bf e} \hat {\bf C}^{\dagger}}{ {\bf e} \hat {\bf C}^{\dagger} {\bf e}^T}\, . 
\end{eqnarray} 

The data covariance matrix $\hat {\bf C}$ follows, $\hat {\bf C} = \hat \sigma^{2}_c {\bf e}^T{\bf e} + \hat {\bf C}_{fc} + {\bf C}_f$
where $\hat \sigma^{2}_c$ represents the variance of the CMB component which is independent on the frequency, $\hat {\bf C}_{fc}$ is an $n \times n$ matrix 
denoting the chance correlation between the CMB and all foreground components for a given realization of CMB (e.g., pure 
CMB signal in our Universe) and finally ${\bf C}_f$ is the 
$n \times n$ foreground covariance matrix\footnote{In this discussions we have assumed that the detector noise contribution
is small compared to the foreground or CMB signal. This is the case for WMAP and Planck temperature observation
over the large scales of the sky. We emphasize that  we do not require detector noise to be completely absent, we only assume that 
the data signal dominated. Accordingly, we interpret ${\bf C}_f$ to contain a small amount of detector noise also}.  
Following~\cite{Saha2008} we note that, ${\bf e}^T \in \mathcal C(\hat {\bf C}_{fc} +  {\bf C}_f)$
so that the generalized Sherman-Morrison formula for Moore-Penrose generalized inverse of rank one update becomes,  
\begin{eqnarray}
\hat {\bf C}^{\dagger} = \hat {\bf A}^{\dagger} - \frac{1}{\lambda} {\bf fg}^{T} \, ,  
\label{GSM}
\end{eqnarray} 
where $\hat {\bf A} = \hat {\bf C}_{fc} +  {\bf C}_f$, $\lambda = 1 + {\bf e}\hat {\bf A}^{\dagger}{\bf e}^T$, ${\bf f} = \hat {\bf A}^{\dagger}{\bf e}^T$ 
and ${\bf g} = \hat {\bf A}^{\dagger}{\bf e}^T$. 
Using Eqn.~\ref{GSM} we obtain, 
\begin{eqnarray}
{\bf W} = %\frac{ {\bf e} \hat {\bf C}^{\dagger}}{ {\bf e} \hat {\bf C}^{\dagger} {\bf e}^T} = 
\frac{ {\bf e} \hat {\bf A^{\dagger}}}{ {\bf e} \hat {\bf A}^{\dagger} {\bf e}^T}\, . 
\label{weights1}
\end{eqnarray}    
Using Eqn.~\ref{weights1} we conclude that the weights are independent on the exact level of CMB variance $\hat \sigma^2_c$ for the 
particular random realization. This is expected since the weights in the usual ILC method, in principle,  should only be determined 
by the foregrounds as long as CMB follows blackbody distribution. One may interpret  Eqn.~\ref{weights1}
as the usual ILC weights minimizing the part of the variance in the cleaned map that arise due to 
CMB-foreground chance correlation and foreground components.  Since $\hat {\bf A} = \hat {\bf C}_{fc} + {\bf C}_f $, 
we see from Eqn.~\ref{weights1} that in practice the 
ILC weights not only depend on foreground covariance matrix ${\bf C}_f$ but also they depend upon the CMB-foreground 
chance correlation matrix $\hat {\bf C}_{fc}$.  
What will happen if in Eqn.~\ref{weights1} we could replace $\hat {\bf A}$ by ${\bf C}_f$? We note that 
such a choice is not possible for analysis of the real data since the covariance matrix for the foregrounds 
is not known exactly a priori. However, in Monte Carlo simulations we can always assume that ${\bf C}_f$ is known.  
This will be the situation when weights are not affected by the chance-correlation matrix. If we know the true
foreground covariance matrix accurately, in usual ILC procedure one will just minimize the part of the 
variance in the cleaned map that arise due to foreground  components. Clearly this is $\sigma^2_f = {\bf W }{\bf  C}_f {\bf W}^T$. 
Minimizing $\sigma^2_f $  subject to the constraint CMB is preserved gives, 
\begin{eqnarray}
{\bf W} = \frac{ {\bf e} {\bf C}^{\dagger}_f}{ {\bf e} {\bf C}^{\dagger}_f {\bf e}^T}\, . 
\label{weights2}
\end{eqnarray} 
We perform detailed Monte Carlo simulations of foreground minimization at low 
resolution  following usual ILC method, with simulated WMAP and Planck observations  to 
investigate the difference in the cleaned maps obtained by two  different ways. First, the weights 
are determined following Eqn.~\ref{weights1} and second, they are determined by Eqn.~\ref{weights2}. 
In the first case we recover results that are similar to those shown in Figs.~\ref{ErrOldILC} and~\ref{ErrOldILC1}. 
This implies  in presence of CMB-foregrounds chance correlations usual ILC method perform a poor foreground 
subtraction on large scales on the sky. In the second case,  when the chance correlation matrix is absent  
the method  performs foreground removal  very well. The standard deviation map computed from the difference 
of cleaned CMB maps and the corresponding input CMB maps, for this case,  is shown in Fig.~\ref{ErrCf}. The standard deviation map is consistent with a  detector 
noise pattern without any visible signature of residual foregrounds. The mean pixel variance  of this map  is only $0.14 \mu K^2$
indicating greatly improved foreground subtraction compared to the case when CMB-foreground chance 
correlation is present. We reemphasize that, although, we can use Eqn.~\ref{weights2} for the case of Monte 
Carlo simulations where the input foreground models are known, in practice, we can not use this equation 
to estimate ILC weights since the foreground covariance matrix ${\bf C}_f$  is unknown for the 
observed sky.  We use Eqns.~\ref{weights2} and~\ref{weights1} in Monte Carlo simulations to establish that the 
CMB-foreground chance correlations cause significant residuals in usual ILC method. The global ILC method that 
propose to use  CMB covariance information, thus, becomes greatly 
beneficial method, improving performance of  usual ILC method without any need to know ${\bf C}_f$.    

Apart from comparing the pixel reconstruction error maps (e.g.,  Figs.~\ref{ErrOldILC} and~\ref{ErrCf}) or the 
power spectra of cleaned maps there is another way in which we see that using the theoretical CMB covariance 
matrix helps to greatly improve usual ILC results. In Fig.~\ref{W} ($x,y$) coordinates of any  blue  point are 
given respectively by value of weight for a particular frequency band obtained using the usual ILC method  and the corresponding 
value of the weight using Eqn.~\ref{weights2} while cleaning a given set of input frequency maps. The $y$-coordinate 
of the  yellow points are same as the blue point for the same set of input frequency maps, however the $x$-coordinate 
of yellow points represents weights for the global ILC method using information about the theoretical CMB covariance 
matrix. The blue points show significantly  larger dispersion along  the horizontal axes  for all frequency bands
compared to the corresponding dispersion of yellow points. The new method of this work efficiently reduces the larger dispersion 
of weights of usual ILC method and produces better foreground minimized CMB maps at low resolution. The $y$-coordinates 
 of all points of this figure show some level of fluctuations, even if  we use 
Eqn.~\ref{weights2}  to estimate the weights that represent the $y$-coordinates. This is because apart from the foregrounds 
${\bf C}_f$ contain a small level of detector noise. The $x$-coordinate of vertical axis of each plot show the value of the
weight when no detector noise is present in ${\bf C}_f$. Each of these values remains same for different  Monte Carlo
simulations and represent weights that will be necessary to remove foregrounds in an ideal noise-less experiment.  
We finally note that using CMB theoretical covariance matrix in Eqn.~\ref{dispersion0} we efficiently suppress CMB large angle covariances which 
leads to significantly smaller dispersion of weights because of smaller CMB-foreground chance correlation. The small dispersion of our weights results in 
a greatly improved foreground minimization than the usual ILC method on large scales of the sky.

\section{Discussions \& Conclusion}
\label{Conclusion}
We have developed a new ILC method for foreground minimization in  pixel space for application on large 
angular scales on the sky using prior information about theoretical CMB covariance matrix. We apply the 
methodology on low resolution  WMAP and  Planck frequency maps 
and show that the cleaned CMB temperature anisotropy map obtained by us match very well with those obtained 
by other science groups of Planck and WMAP. {\it This shows that results of CMB maps and its power spectrum are 
robust with respective to a variety of analysis pipeline.} We validate the methodology of our foreground 
removal by detailed Monte Carlo simulations. Usage of this new approach has several benefits over  naive 
application of usual ILC approach in pixel space over large scales of the sky. 

\begin{enumerate}
\item First, the new approach generates cleaned CMB map that has significantly 
lower reconstruction error due to foreground residuals.  The power spectrum from the cleaned map also has the 
lower reconstruction error for our case, the standard deviations of CMB angular power spectrum estimated 
from the Monte-Carlo simulations agree with those estimated from the cosmic variance alone. 
\item Second, the  
CMB angular power spectrum obtained from our cleaned maps does not have any  visible signature of negative 
bias at the low multipole region, which is seen to be present for pixel space application of usual 
ILC method over large scales on the sky. Such negative  bias is also reported in harmonic space ILC 
method by~\cite{Saha2006} and its property and origin were investigated in detail by~\cite{Saha2008}.    
The negative bias arise  due a chance correlation between CMB and foreground components on a particular
realizations of the sky. Using inverse weight of CMB theoretical covariance matrix in Eqn.~\ref{dispersion0} we 
effectively get rid of such chance correlations and the as well as the resulting negative biases in the 
cleaned CMB angular power spectrum at low multipoles. 
\end{enumerate} 

The new method complements the usual ILC approach in pixel space which so far has been applied 
on high resolution maps by incorporating local information available from input frequency maps to better 
remove foregrounds, the spectral property of which vary with the sky positions. On the very large
scales the spectral properties of foregrounds are expected to vary by small amount over the 
entire sky. We show that, on the large scale it is sufficient to perform ILC foreground 
removal by dividing the sky  merely into two regions, provided we use the prior information 
available from CMB covariance matrix globally on the sky. Although we have assumed a theoretical
CMB covariance matrix consistent with assumption of statistical isotropy of CMB in Eqn.~\ref{theory_cov}, 
in principle, one can also use a covariance matrix in our method which is not statistically isotropic. This 
brings about a possibility to open up a new avenue to incorporate such additional information in our method 
which may be a signature of non trivial primordial power spectrum~\citep{Ghosh2016,Contreras2017}. Taking into account the global 
nature of our low resolution analysis and local nature of high resolution analysis of usual ILC 
method, we now consider pixel space ILC method in a general  perspective that incorporates
a very comprehensive duality in its nature.  
We hope that our method will be useful to analyze low resolution polarization maps from Planck or 
future generation CMB missions.

 We use publicly available HEALPix~\cite{Gorski2005} package available from 
http://healpix.sourceforge.net for some of the analysis of this work. 
We acknowledge the use of Planck Legacy Archive (PLA) and the Legacy Archive for Microwave Background 
Data Analysis (LAMBDA). LAMBDA is a part of the High Energy Astrophysics Science Archive Center (HEASARC). 
HEASARC/LAMBDA is supported by the Astrophysics Science Division at the NASA Goddard Space Flight
Center.

%%\bibliography{ms}
%%\bibliographystyle{apj}

\end{document}